%% file: ijcds_master_doc.tex
\documentclass[journal,twoside,hidelinks,a4paper]{IEEEtran}

\input{ijcds_header.tex}

\begin{document}
\bstctlcite{IEEEexample:BSTcontrol}
\input{ijcds_body.tex}
\bibliographystyle{IEEEtran}
\bibliography{references}
\input{ijcds_biographies.tex}
\end{document}

%% file: ijcds_header.tex
\newcommand{\subparagraph}{}

\usepackage{color, colortbl}
\definecolor{LightCyan}{rgb}{0.88,1,1}
\definecolor{ivory}{rgb}{1.0, 1.0, 0.94}
\definecolor{lightyellow}{rgb}{1.0, 1.0, 0.88}
\definecolor{piggypink}{rgb}{0.99, 0.87, 0.9}

\usepackage{pifont} 

\usepackage[linesnumbered,ruled,vlined,commentsnumbered]{algorithm2e}
\usepackage{amsmath}
\usepackage{rotating}
\usepackage{emptypage,titling,titlesec,widetext,textcomp,relsize,parskip}
\usepackage[utf8]{inputenc}
\usepackage[margin=10pt,font=footnotesize,labelsep=period]{caption}
\usepackage{tabulary,longtable,makecell,multirow,color,colortbl}
\usepackage[colorlinks=false,linkcolor=black,urlcolor=gray,citecolor=black]{hyperref}
\usepackage[usenames,dvipsnames,svgnames,table]{xcolor}
\usepackage{authblk,lipsum,txfonts,float,morefloats,gensymb,graphicx,setspace,fancyhdr,bibunits,booktabs,array,marvosym,xhfill}
\usepackage[normalem]{ulem}
\usepackage[export]{adjustbox}

\usepackage{fontawesome}
\usepackage{placeins}

\usepackage[includeheadfoot,paperwidth=8.5in,paperheight=11in,top=1cm,bottom=2cm,left=2cm,right=2cm,footskip=.25in,headheight=40pt,headsep=12pt,heightrounded]{geometry}

\newcommand{\etal}{\textit{et}~\textit{al.}~}
\newcommand{\ie}{\textit{i.e.,}~}
\newcommand{\eg}{\textit{e.g.,}~}
\newcommand{\resp}{\textit{resp.}~}
\newcommand{\etc}{\textit{etc.}~}
\newcommand{\cf}{\textit{cf.}~}
\newcommand{\aka}{\textit{aka.}~}

\newcommand{\squeezeup}{\vspace{-3.5mm}}
\newcommand{\xfill}[2][1ex]{{\dimen0=#2\advance\dimen0 by #1\leaders\hrule height \dimen0 depth -#1\hfill}}
\newcommand{\xfilll}[2][1ex]{\dimen0=#2\advance\dimen0 by #1\leaders\hrule height \dimen0 depth -#1\hfill}

\renewcommand{\figurename}{Figure}

\titlespacing\section{0pt}{8pt plus 4pt minus 2pt}{2pt plus 2pt minus 2pt}
\titlespacing\subsection{0pt}{8pt plus 4pt minus 2pt}{2pt plus 2pt minus 2pt}
\titlespacing\subsubsection{0pt}{8pt plus 4pt minus 2pt}{2pt plus 2pt minus 2pt}

\newcommand{\subtitle}[1]{\posttitle{\par\end{center}\begin{center}\bfseries\Large #1\end{center}\vskip1.2em}}
\newcommand\myabstract[2][.8]{\renewcommand\maketitlehookd{\mbox{}\medskip\par\centering\begin{minipage}{#1\textwidth}\small#2\end{minipage}}}

\definecolor{myred}{RGB}{189, 3, 34}
\date{}

\title{\vspace{-1cm}{\includegraphics[height=0.5in,keepaspectratio=true]{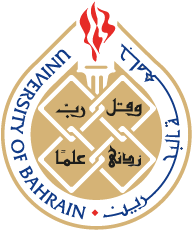}\hfill\parbox{15cm}{\raggedleft\sublargesize{\color{myred}\bfseries{International Journal of Computing and Digital Systems\\}}\sublargesize{ISSN (2210-142X)\\Int. J. Com. Dig. Sys. 15, No.1 (Jan-24)\\\vspace{3mm} \url{http://dx.doi.org/10.12785/ijcds/150118}}}\\\vspace{1cm}{\bfseries{\fontsize{18}{21.6}BushraDBR:}}}}

\subtitle{An Automatic Approach to Retrieving Duplicate Bug Reports}

\author[1]{\bfseries Ra'Fat Al-Msie'deen}

\affil[1]{\normalfont\textit{Department of Software Engineering, Faculty of IT, Mutah University, Mutah 61710, Karak, Jordan}}
\affil[ ]{\vspace{0pt}}
\affil[ ]{Received 27 Feb. 2023, Revised 3 Jan. 2024, Accepted 6 Jan. 2024, Published 15 Jan. 2024}
\affil[ ]{\vspace{0pt}}

\fancypagestyle{firstpage}{%
    \fancyhf{}
    \fancyhead[RO]{{\normalfont{Int. J. Com. Dig. Sys. 15, No.1, 221-238 (Jan-24)\hspace{4mm}}}\includegraphics[height=0.5in, keepaspectratio=true]{ub_logo}\hspace{3mm}\bfseries{\thepage}}
    \fancyhead[LE]{\bfseries{\thepage}\hspace{3mm}\includegraphics[height=0.5in,keepaspectratio=true]{ub_logo}\hspace{3mm}\normalfont\textit{Ra'Fat Al-Msie'deen: BushraDBR — An Automatic Approach to Retrieving Duplicate Bug Reports.}}
    \fancyfoot[R]{\vspace{5mm}\textsl{\color{gray}{\textsf{\url{https://journal.uob.edu.bh/}}}}}}

\fancypagestyle{plain}{%
    \fancyhf{}
    \fancyhead[LR]{\vspace{2.5cm}}
    \fancyfoot[L]{\vspace{0mm}{\small\textit{E-mail address: rafatalmsiedeen@mutah.edu.jo}}}
    \fancyfoot[R]{\vspace{5mm}\textsl{\color{gray}{\textsf{\url{https://journal.uob.edu.bh/}}}}}}

%% file: ijcds_body.tex
\myabstract[1]{\vspace{-1cm}\hrule\vspace{2mm}\textbf{Abstract:} A Bug Tracking System (BTS), such as Bugzilla, is generally utilized to track submitted Bug Reports (BRs) for a particular software system. Duplicate Bug Report (DBR) retrieval is the process of obtaining a DBR in the BTS. This process is important to avoid needless work from engineers on DBRs. To prevent wasting engineer resources, such as effort and time, on previously submitted (or duplicate) BRs, it is essential to find and retrieve DBRs as soon as they are submitted by software users. Thus, this paper proposes an automatic approach (called BushraDBR) that aims to assist an engineer (called a triager) to retrieve DBRs and stop the duplicates before they start. Where BushraDBR stands for Bushra Duplicate Bug Reports retrieval process. Therefore, when a new BR is sent to the Bug Repository (BRE), an engineer checks whether it is a duplicate of an existing BR in BRE or not via BushraDBR approach. If it is, the engineer marks it as DBR, and the BR is excluded from consideration for any additional work; otherwise, the BR is added to the BRE. BushraDBR approach relies on Textual Similarity (TS) between the newly submitted BR and the rest of the BRs in BRE to retrieve DBRs. BushraDBR exploits unstructured data from BRs to apply Information Retrieval (IR) methods in an efficient way. BushraDBR approach uses two techniques to retrieve DBRs: Latent Semantic Indexing (LSI) and Formal Concept Analysis (FCA). The originality of BushraDBR is to stop DBRs before they occur by comparing the newly reported BR with the rest of the BRs in the BTS, thus saving time and effort during the Software Maintenance (SM) process. BushraDBR also uniquely retrieves DBR through the use of LSI and FCA techniques. BushraDBR approach had been validated and evaluated on several publicly available data sets from Bugzilla. Experiments show the ability of BushraDBR approach to retrieve DBRs in an efficient and accurate manner.
\newline
\newline
\textbf{Keywords:} Software engineering, Software maintenance, Duplicate bug report retrieval, Formal concept analysis, Latent semantic indexing, Bug tracking system, Bug report. \vspace{1mm} \hrule}

\maketitle

\section{Introduction}
\label{sec:introduction_and_overview}

A software bug is a fault in its code, made by software coders, that prevents software from running properly. Software bugs are reported as BRs to a BTS during SM. Hundreds of BRs are submitted every day for large and complex software products (\eg Mozilla and Eclipse). Duplicated BRs arise when multiple users report multiple BRs for the same software bug \cite{DBLP:AnvikHM06}. Because of the asynchronous nature of the BR submission process, conventional BTSs (\eg Bugzilla) cannot avoid DBRs. Thus, some BRs remain duplicates of one another in BTSs \cite{DBLP:ZouLCXFX20}. DBRs cause a main overhead in SM process since they usually cost valuable development time, effort, cost, and resources \cite{DBLP:JalbertW08}. DBR detection (\aka BR de-duplication) is a hot topic in the software engineering field \cite{DBLP:HindleO19} \cite{DBLPZhangHVIXTLJ23}.

Software bugs are inevitable in software products due to their complexity. One of the most significant activities during SM is bug fixing. Software failures affected many stockholders and caused financial losses. Therefore, knowing how to correctly fix as many bugs as possible is a big help in the development of software products \cite{DBLP:ZouLCXFX20}. Nowadays, BRs have been playing an essential role in bug fixing since they offer certain details (\eg bug summary and description) to help software engineers locate and fix specific defects in software code \cite{DBLP:ZimmermannPBJSW10}.

Large software products (\eg Eclipse) continuously receive several BRs, especially after main releases when users report new bugs \cite{OscarChaparro2019}. In order to confirm and assign the reported bug to an engineer for fixing, it is important to check if this bug has been submitted before \cite{DBLP:DavidsonMJ11}. When DBRs are found, BRs are marked as a result, thus preventing possibly redundant work and the cost of bug fixing \cite{DBLP:BettenburgPZK08}. As soon as the number of BRs is huge, detecting (or retrieving) DBRs becomes a time-consuming, costly, and error-prone task. Therefore, this paper suggests an automatic approach that aims to save time, cost, and effort and increase the accuracy of the DBR detection process. \textit{BushraDBR} is an open-source approach for engineers to help them retrieve possible DBRs before assigning them to software developers.

The bug reporting activity is an essential part of the SM process. Nowadays, BTS is employed by software users to maintain the track record of a software bug that is reported throughout the usage of a specific software system \cite{Palvika}. The key input for any BTS is BRs. BTS upholds the Master BRs set (MRs). Commonly, natural language is employed to write BRs. The same BR can be written in several ways by the software user who reports the bug. It is because the vocabulary differs across software users based on their level of technical background. The content of BR is later examined by a specialist who has a complete understanding of the software known as triager \cite{DBLP:KukkarMKNBK20}.

Triager has two key responsibilities. Where triager converts the wording of BR into more technical language for better comprehension by the software developers. Also, triager carries out the search activity in MRs (or BRE) for possible DBRs that have a similar (or common) signature. Furthermore, if the New BR (Nr) is non-duplicate, then it is joined to MRs; otherwise, it is deemed a DBR. On the other hand, the filtering task of DBR requires a large amount of time, manual effort, and a comprehensive understanding of BRs \cite{DBLP:AngelShiny}.

SM activity is referred to as the modification process of a software system after its delivery to end users in order to correct software errors (or bugs) \cite{IEEE}. This activity aims at improving software properties (\eg performance) \cite{ALmsiedeen12202111}. Sometimes, software maintenance is important to adapt software products to a modified workplace (\ie environment). Bug triaging is a significant, tedious, and time-consuming task of SM activity \cite{DBLPGuptaIF22}.

\begin{figure}[!htb]
	\center
	\includegraphics[width=0.60\columnwidth]{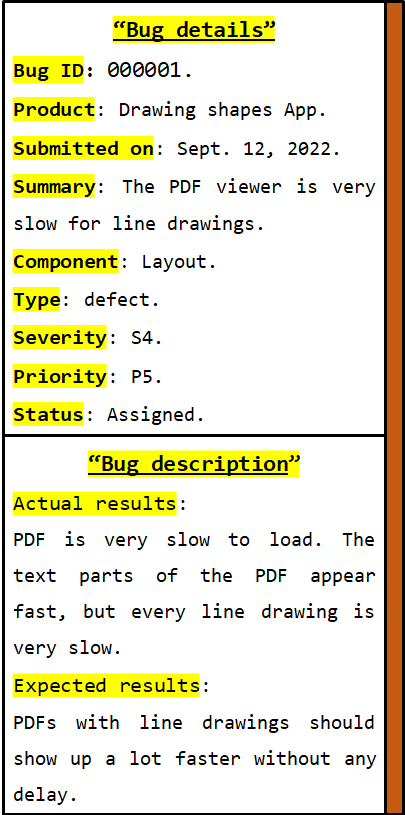}
	\caption{An example of a BR from drawing shapes application.}
	\label{Fig:BRexample}
\end{figure}

DBRs are assigned to several developers for fixing the bug, which wastes developer effort and time. Thus, automating the DBR retrieval procedure is extremely valuable. It decreases the developer's time, effort, and cost. Also, the decrease in manual effort improves developers' productivity. It also reduces the cost of SM activity \cite{DBLP:AnvikHM05}. Each BR has two kinds of information (\ie structured and unstructured). Structured information includes specific information regarding the bug (\eg product, component, and severity of the bug). While the unstructured information includes a natural language description of the bug (\eg bug summary and description). The description of a software bug may be long or short \cite{DBLP:WangZXAS08}. A BR includes numerous pieces of information that are related to a specific bug or issue. Figure \ref{Fig:BRexample} gives an example of a BR from the \textit{Drawing Shapes Application} (DSA) \cite{ALmsiedeen125022} \cite{ALmsiedeen154} \cite{RaFatRT}. DSA lets the end user draw numerous types of shapes, like lines, rectangles, and ovals \cite{RaFatImpact2018} \cite{FatMsiedeenOODocumentation}.

It has been seen that often a BR reported is a duplicate, which results in large DBRs in a BTS \cite{DBLP:AnvikHM05}. When multiple users report BRs for the same trouble, these reports are known as DBRs. As demonstrated in existing studies, the ratio of DBRs can be up to 30\% \cite{DBLP:JalbertW08}. DBRs lead to a state where the same bugs are sent to several engineers who reproduce and resolve the bug for the same reason, which is considered a waste of time, effort, and cost \cite{LiKang}. DBRs are a big problem for Quality Assurance (QA) engineers, triagers, testers, and developers since they stimulate additional work to resolve the problem \cite{DBLP:RakhaSH16}. In this paper, DBR retrieval is the process of querying textually similar BRs in order to group BRs that report the same trouble (or issue) in one cluster.

Table \ref{tab:DBRsBugzilla} displays a pair of DBRs from Bugzilla \cite{Bugzilla}. Please note that the two BRs shown in Table \ref{tab:DBRsBugzilla} belong to the same component, \textit{CSS parsing and computation} \cite{CSSPCCom}, from the \textit{core} product \cite{CoreProduct}. Readers can observe that they are textually similar and belong to the same product (\ie core) from Mozilla.

\begin{table}[!htp]
 \center
 \caption{A pair of DBRs from Bugzilla (\ie core product).}
 \begin{tabular}{|m{1.3em}|m{8em}|m{11em}|}	
 \hline	
 \rowcolor{LightCyan}
Bug ID                     &
A Short Summary of the Bug &
Bug Description            \\\hline
\begin{sideways} 671128 \cite{671128} \end{sideways}  &		
"Assertion: container didn't take ownership: 'Not Reached'" &
"Assertion: container didn't take ownership: 'Not Reached', file layout/style/StyleRule.cpp, line, ... \textit{etc.}" \\\hline
\begin{sideways} 803372 \cite{803372} \end{sideways} &	
"Assertion: container didn't take ownership" mutating a deleted, matched CSS rule"&
"Assertion: container didn't take ownership: 'Not Reached', file layout/style/StyleRule.cpp, line, ... \textit{etc.}" \\\hline
    \end{tabular}
 \label{tab:DBRsBugzilla}
\end{table}

In this work, reports that are textually similar to each other are called DBRs. TS between BRs is good evidence that they describe the same (or similar) issue. Frequently, DBRs are reported by multiple software users. These users come from different backgrounds and use different vocabularies to describe the same (or similar) software bug.

The important role of the triager is to check the reported BRs for any possible duplicates before sending them to the BRE. A manual check of submitted BRs is a difficult task due to the huge number of BRs reported every day. On the other hand, retrieving DBR after storing it in the BRE is a tedious and expensive task for software developers. So, the process of stopping DBRs before they start is important and very useful. Thus, current studies have suggested numerous approaches to retrieving and detecting DBRs from BRE (\cf Section \ref{sec:related_work}). The majority of existing approaches detect DBRs within a single data set that contains all software bugs (\resp BRE, BTS, or MRs).

The novelty of this paper is that it proposes a textual-based approach (\ie an IR-based solution) to stopping DBRs before they start. BushraDBR prevents DBRs by continuously checking the recently reported or submitted BR against the BRs stored in the BTS. In the event that the submitted BR is textually similar to any of the BRs inside the BTS, the triager will exclude this BR from any further work and not include it in the BTS.

\begin{figure}[!htb]
	\center
	\includegraphics[width=0.90\columnwidth]{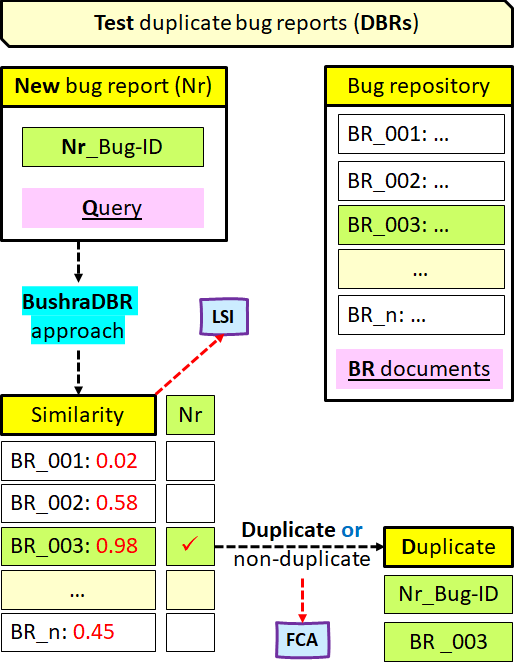}
	\caption{The workflow for retrieving DBRs via BushraDBR approach.}
	\label{Fig:DBReXAM}
\end{figure}

Figure \ref{Fig:DBReXAM} gives an illustrative example regarding BushraDBR approach workflow. BRE contains a collection of BRs (\ie BR\_001, BR\_002, BR\_003, ..., BR\_n). BushraDBR approach uses the newly submitted BR as a query (\ie Nr\_Bug-ID). Then, it calculates the TS between query and BR documents based on LSI \cite{SalmanSDA12} \cite{AlMsiedeenHSUV14}. Thereafter, it retrieves DBRs (if any) using FCA \cite{ALmsiedeenahmad} \cite{FatMsiedeenFMbOOK}. In Figure \ref{Fig:DBReXAM}, the test shows that the query document is associated with a BR document with the ID "BR\_003", where the similarity score equals \textit{0.98} (\ie the highest similarity score). Thus, in this example, BushraDBR makes a successful retrieval of DBRs.

BushraDBR approach uses LSI and FCA to retrieve DBRs. The interested reader can get further information about LSI and FCA techniques from several previous studies \cite{ALMsiUVICSR} \cite{DBLP:Al-MsiedeenHSUV14} \cite{DBLPAlMsiedeenSHUV13} \cite{AlMsieDeen1445} \cite{Doc.SympAl-Msiedeen1} \cite{bookAl-Msiedeen-4}. The suggested approach was validated and evaluated on several data sets from Bugzilla (\cf Section \ref{sec:Exper.}). The results of the experiments show the ability of BushraDBR approach to retrieve DBRs when they exist.

The rest of the paper is structured as follows: A mini-systematic survey regarding DBRs detection and retrieval is presented in Section \ref{sec:related_work}. Section \ref{sec:Approach} details the DBR retrieval process step by step. Experimental results are presented in Section \ref{sec:Exper.}. Finally, Section \ref{sec:conclusion} concludes the paper and talks about the future work of BushraDBR.

\section{A mini-systematic survey about DBR detection and retrieval} \label{sec:related_work}

Numerous approaches have been suggested in the literature to assist in the automatic detection of DBRs. This section offers previous approaches relevant to BushraDBR contributions.

Retrieving DBRs is a time-consuming and boring task because of the various writing styles of the huge number of submitted BRs to BTSs. Therefore, there is a need to propose BushraDBR approach in order to automate the process of DBR retrieval and avert manual effort and analysis. The main idea is to utilize BushraDBR approach for validating whether the newly submitted BR (\ie Nr) is duplicate or non-duplicate. Over the last few years, numerous studies have been suggested on the automatic retrieval and detection of DBRs. The main details of these studies are offered below:

To retrieve DBRs, Wang \etal \cite{DBLP:WangZXAS08} suggested an approach to detecting DBRs. They use natural language and execution information to obtain and determine the TS in MRs (or BRE). The experimental findings showed that the suggested approach can detect 67\%-93\% of DBRs in MRs of the Firefox project, compared to 43\%-72\% by utilizing only natural language information.

Sun \etal \cite{DBLP:SunLWJK10} proposed an approach to detecting DBRs by constructing a discriminative model that detects if two BRs are duplicates of each other or not. Their model reports a score on the possibility of X and Y being duplicates. Then, this score is utilized to retrieve related BRs from a BRE for user examination. On three BREs from Firefox, Eclipse, and OpenOffice, they have studied the feasibility of their method.

Jalbert and Weimer \cite{DBLP:JalbertW08} suggested an approach that automatically classifies DBRs as they arrive at BTS. The main goal of their approach is to save developers' time. This work utilized textual semantics, graph clustering, and surface features to find DBRs. Their technique is suggested to decrease BR triage costs by discovering DBRs as they are submitted. Thus, they built a classifier for the arriving BRs to discover BR duplicates.

Sun \etal \cite{DBLP:SunLKJ11} suggested an approach based on \textit{BM25F} to retrieve DBRs, where \textit{BM25F} is an efficient technique for document similarity measurement. In addition to the textual information of BRs (\ie summary and description), additional information from BRs (\eg product, component, priority, \textit{etc.}) is employed to retrieve DBRs. The authors assess their approach by generating a list of candidate DBRs for every BR indicated as \emph{duplicate} by the bug triager, to determine whether or not the correct duplicate is a candidate. On three MRs from Mozilla, Eclipse, and OpenOffice, they applied their method. They achieved DBR detection improvement with $17-23\%$ in mean average precision and $10-27\%$ in recall rate@k (1 $\leq$ k $\leq$ 20).

Hindle and Onuczko \cite{DBLP:HindleO19} suggested the continuously querying method, which assists software users in finding DBRs as they type in their bug report. Their approach attempts to prevent DBRs before they start by continuously querying. Their approach has the ability to prevent duplicates before they happen in 42\% of cases by building a simple IR model using TF-IDF and cosine distance to find similar BRs from their prefixed queries.

Thung \etal \cite{DBLP:ThungKL14} have proposed a tool named \textit{DupFinder} that finds the DBRs. DupFinder mines texts from the summary and detailed description of a new BR and BRs present in a BTS. Also, it uses the Vector Space Model (VSM) to measure the similarity of BRs and provides the software triager with a list of possible DBRs based on the TS of these BRs with the new BR.

Kukkar \etal \cite{DBLP:KukkarMKNBK20} suggested an automatic approach for DBRs detection and classification by using the Deep Learning (DL) technique. The suggested approach has three components, which are: preprocessing, the DL model, and DBR detection and classification. Their approach utilized a Convolutional Neural Network (CNN) based DL model to obtain the appropriate feature. These appropriate features are utilized to define the similar features of BRs. Thus, the BRs similarity is computed based on these similar features. Their study calculated the similarity value of two BRs. Then, BRs are categorized as duplicate or non-duplicate based on the similarity values. The suggested approach applied to several available data sets, like Mozilla, NetBeans, and Gnome. The results are assessed using different metrics, such as an F-measure, recall@k, accuracy, precision, and recall. The accuracy rate of the suggested approach is between 85\% and 99\%, and its recall@k rate is between 79\% and 94\%, according to experimental results. 
The results of BurshraDBR approach are evaluated using different metrics, such as recall, precision, and F-measure. The recall rate of BurshraDBR is 100\%, and its precision rate is 100\%, according to experimental results, which demonstrate that BurshraDBR approach performs better than current approaches.  

He \etal \cite{DBLPHeXY0L20} have suggested a method to detect DBRs in pairs by employing CNN. They suggested creating a single representation for each pair of BRs via the Dual-Channel Matrix (DCM). DCM is fed to CNN to discover correlated semantic links between BRs. Then, the suggested method determines whether a pair of BRs is duplicate or not by using association features. They evaluate the suggested method on three data sets from Open Office, Eclipse, and NetBeans projects. Findings show that their method achieves excellent accuracy.

In \cite{DBLPRunesonAN07}, Runeson \etal have suggested a method to detect DBRs by employing Natural Language Processing (NLP). Findings demonstrate that their method can find 2/3 of DBRs by means of NLP techniques. In their work, they have described the five processing stages in NLP based on Manning and Sch{\"{u}}tze \cite{DBLP0001548}, which are: tokenization (or word splitting) \cite{AuLabRafat}, stemming (root of words) \cite{RaFat2019TagClouds}, English stop words removal, representation of the vector space, and, at last, calculating similarity values. From their work, BurshraDBR's approach uses some steps in the pre-processing of BRs, such as tokenization, stemming, and removing English stop words.

In the literature, there are numerous works assessing DBR detection (or retrieval) approaches \cite{DBLPRakhaBH18}, \cite{DBLPRakhaBH182}. Rakha \etal \cite{DBLPRakhaBH182} analyzed and assessed the changes between DBRs before and after the offering of the \textit{Just-In-Time} (JIT) DBR recommendation feature in Bugzilla. The JIT duplicate retrieval feature was introduced in 2011 for Bugzilla $4.0$ \cite{Bugzilla4}. The authors have discovered that DBRs after 2011 (the period between 2012 and 2015) are less textually similar than duplicate BRs before the activation of the JIT feature in 2011.

In \cite{Neysiani2020}, Neysiani and Babamir proposed a study aimed at assessing the best DBR detection (or retrieval) approaches. They analyzed both \textit{IR-based} and \textit{Machine Learning} (ML) approaches. The study has proven that \textit{ML-based} approaches are more effective and efficient than IR-based approaches. The research was assessed just in the Android BRE.

\begin{figure}[!htb]
	\center
	\includegraphics[width=0.93\columnwidth]{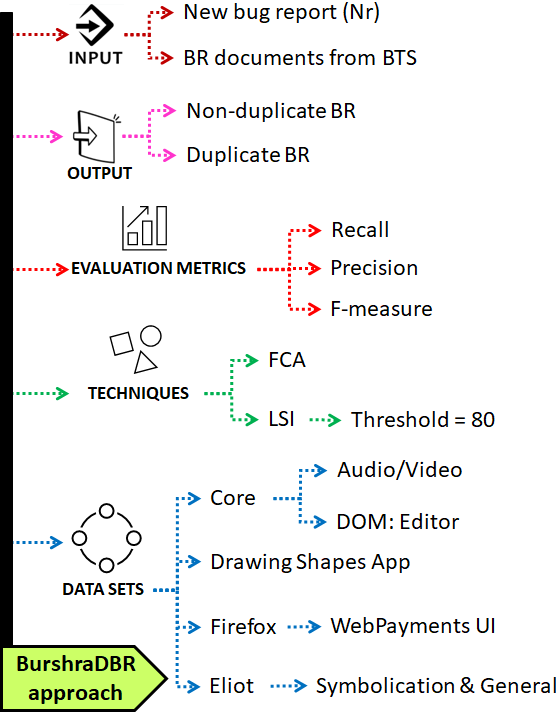}
	\caption{The \textit{key elements} of BurshraDBR approach.}
	\label{Fig:keyElements}
\end{figure}

This section gives the most recent and relevant studies regarding BurshraDBR's contributions. Figure \ref{Fig:keyElements} illustrates the key elements of BurshraDBR's approach. The author summarizes the suggested approach according to the following basic elements: inputs (\ie BRs and Nr), outputs (\ie duplicate or non-duplicate BR), used techniques (\ie LSI and FCA), evaluation metrics (\eg recall), and used data sets (\eg Core, Firefox, and Eliot products from Bugzilla).

In this study, the author extracts all the essential parts for BurshraDBR from each BR, such as bug summary and description (\ie unstructured information). On the other hand, some approaches utilize structured information such as bug ID, product, and component. Table \ref{tab:Info.} shows the type of information that each approach utilizes in order to retrieve DBRs from MRs or BRE.

\begin{table}[!htb]
	\center
	\caption{Structured and unstructured information that is leveraged by the selected approaches (\ie survey).}
	\begin{tabular}{|l|c|c|c|c|c|c|}	\hline	
	\rowcolor{LightCyan}
	Reference   & \multicolumn{4}{c|}{Structured Info} & \multicolumn{2}{c|}{Unstructur.} \\\cline{2-7}
	&
	\begin{sideways} Bug ID      \end{sideways} &
	\begin{sideways} Product     \end{sideways} & 
	\begin{sideways} Component   \end{sideways} & 
	\begin{sideways} Type        \end{sideways} & 
	\begin{sideways} Summary     \end{sideways} &
	\begin{sideways} Description \color{lightyellow}.\end{sideways} \\\hline	
Wang \etal \cite{DBLP:WangZXAS08}       &         &        &        &        &$\times$ &$\times$  \\\hline
Sun \etal \cite{DBLP:SunLWJK10}         &         &        &        &        &$\times$ &$\times$  \\\hline
Jalbert \& Weimer \cite{DBLP:JalbertW08}&         &        &        &        &$\times$ &$\times$  \\\hline
Sun \etal \cite{DBLP:SunLKJ11}          &         &$\times$&$\times$&$\times$&$\times$ &$\times$  \\\hline
Hindle \& Onucz. \cite{DBLP:HindleO19}  &         &        &        &        &$\times$ &$\times$  \\\hline
Thung \etal \cite{DBLP:ThungKL14}       &         &        &        &        &$\times$ &$\times$  \\\hline
Kukkar \etal \cite{DBLP:KukkarMKNBK20}  &$\times$ &        &        &        &$\times$ &$\times$  \\\hline
He \etal \cite{DBLPHeXY0L20}            &         &$\times$&$\times$&        &$\times$ &$\times$  \\\hline
Runeson \etal \cite{DBLPRunesonAN07}    &         &$\times$&        &        &$\times$ &$\times$  \\\hline
Rakha \etal \cite{DBLPRakhaBH18}        &$\times$ &        &$\times$&$\times$&$\times$ &$\times$  \\\hline
Neysiani \& Ba. \cite{Neysiani2020}     &         &$\times$&$\times$&$\times$&$\times$ &$\times$  \\\hline
\rowcolor{piggypink}
BushraDBR                               &\ding{51}&        &        &        &\ding{51}&\ding{51} \\\hline
		\end{tabular}
	\label{tab:Info.}
\end{table}

A study and comparison of current studies confirmed that there are no works in the literature that use LSI and FCA to retrieve DBRs. In this work, LSI and FCA techniques are applied in order to retrieve DBRs. BushraDBR exploits the bug's summary and description to construct a BR document. Also, BushraDBR visualizes the retrieved TS scores between BRs.
Table \ref{tab:Info.DS} shows the data sets used in each study from the survey (\ie related work). The mini-systematic survey showed that the following data sets are most commonly used in research papers: Eclipse, OpenOffice, and Mozilla (\cf Table \ref{tab:Info.DS}).

\begin{table}[!htb]
	\center
	\caption{Data sets that are utilized by the selected approaches (\ie survey).}
		\begin{tabular}{|l|c|c|c|c|c|c|c|c|c|c|}	\hline	
			\rowcolor{LightCyan}
		    & \multicolumn{10}{c|}{Bug report data sets} \\\cline{2-11}
			\begin{sideways} Reference                     \end{sideways}&     
			\begin{sideways} Eclipse                       \end{sideways}&
			\begin{sideways} Firefox                       \end{sideways}&
			\begin{sideways} OpenOffice                    \end{sideways}& 
			\begin{sideways} Mozilla                       \end{sideways}&
			\begin{sideways} Android                       \end{sideways}&
			\begin{sideways} Sony Ericsson Mobile \color{lightyellow}- \end{sideways}&
			\begin{sideways} Cyanogenmod                   \end{sideways}&
			\begin{sideways} NetBeans                      \end{sideways}&
			\begin{sideways} Gnome                         \end{sideways}&
			\begin{sideways} K9Mail                        \end{sideways}\\\hline

			\smaller \textbf{Wa\cite{DBLP:WangZXAS08}}    & $\times$&$\times$& &&&&&&& \\\hline
			\smaller \textbf{Su\cite{DBLP:SunLWJK10}}     & $\times$&$\times$&$\times$&&&&&&& \\\hline
			\smaller \textbf{Jalb\cite{DBLP:JalbertW08}}  & &&&$\times$&&& &&& \\\hline
			\smaller \textbf{Su\cite{DBLP:SunLKJ11}}      & $\times$&&$\times$&$\times$&&& &&& \\\hline
			\smaller \textbf{Hin\cite{DBLP:HindleO19}}    & $\times$&&$\times$&$\times$&$\times$&&$\times$&&&$\times$ \\\hline
			\smaller \textbf{Th\cite{DBLP:ThungKL14}}     & &&&$\times$&& & &&& \\\hline
			\smaller \textbf{Ku\cite{DBLP:KukkarMKNBK20}} &$\times$&$\times$&$\times$&$\times$&&& &$\times$&$\times$& \\\hline
			\smaller \textbf{He\cite{DBLPHeXY0L20}}       & $\times$&&$\times$&&& & &$\times$&& \\\hline
			\smaller \textbf{Ru\cite{DBLPRunesonAN07}}    && &&& &$\times$&&&& \\\hline
			\smaller \textbf{Ra\cite{DBLPRakhaBH18}}      &$\times$& &$\times$&$\times$&& & & && \\\hline
			\smaller \textbf{Ne\cite{Neysiani2020}}       &&&&&$\times$ &&& && \\\hline
			\rowcolor{piggypink}
			\smaller{\textbf{Bushra}}                     & &\ding{51}&&\ding{51}&& &&&& \\\hline
		\end{tabular}
	\label{tab:Info.DS}
\end{table}

Related work shows that there are several approaches to retrieving DBRs. The author categorizes those approaches into three categories: IR-based, ML-based, and hybrid solutions. Table \ref{tab:relatedWork} presents the majority of existing approaches that retrieve DBRs. Table \ref{tab:relatedWork} categorizes current studies based on the type of approach used (\ie IR, ML, or hybrid).

\begin{table*}[!htb]
	\center
	\caption{Review and classification of current studies relevant to DBR retrieval approaches.}
		\begin{tabular}{|m{1.0em}|m{10em}|m{35em}|}	\hline	
			\rowcolor{LightCyan}
			 ID & Type of approach & Approaches (or references)  \\\hline	
			 		 
			 01&IR-based approach  & \underline{\textit{BushraDBR}}, Jalbert and Weimer \cite{DBLP:JalbertW08}, Sun \etal \cite{DBLP:SunLKJ11}, Hindle and Onuczko \cite{DBLP:HindleO19}, Thung \etal \cite{DBLP:ThungKL14}, Runeson \etal \cite{DBLPRunesonAN07}, Rakha \etal \cite{DBLPRakhaBH182}, Li \etal \cite{DBLPLiYYWW17}, Li \etal \cite{DBLPLiYWYMW21}, Sureka and Jalote \cite{AshishJaloteSureka}, Wang \etal \cite{DBLP:WangZXAS08}, Rakha \etal \cite{DBLPRakhaBH18}, Banerjee \etal \cite{DBLPBanerjee12}, Banerjee \etal \cite{DBLPBanerj17}. \\\hline
			 
			 02&ML-based approach  & Sun \etal \cite{DBLP:SunLWJK10}, Kukkar \etal \cite{DBLP:KukkarMKNBK20}, He \etal \cite{DBLPHeXY0L20}, Deshmukh \etal \cite{DBLPDeshmukhMPSD17}, Aggarwal \etal \cite{DBLPAggarwalTRHSG17}, Budhiraja \etal \cite{DBLPBudhirajaDRS18}, Budhiraja \etal \cite{DBLPBudhirajaDSR18}, Messaoud \etal \cite{Messaoud}, Wu \etal \cite{DBLPWuSZCRS23}, Panichella \cite{DBLPPanichella19}, Isotani \etal \cite{DBLPIsotaniWFNOS22}, Alipour \etal \cite{DBLPAlipourHS13}. \\\hline
			 
			 03&Hybrid approach (IR \& ML)& Bagal \etal \cite{BagalPrasad}, Tian \etal \cite{DBLPTianSL12}, Neysiani and Babamir \cite{Neysiani2020}, Nguyen \etal \cite{DBLNguyenNNLS12}, Jiang \etal \cite{DBLPJiangSTSW23}, Pasala \etal \cite{PasalaAnjaneyulu}, Feng \etal \cite{DBLPFengSSG13}. \\\hline
		\end{tabular}
	\label{tab:relatedWork}
\end{table*}

\section{The DBR retrieval process: BushraDBR approach}\label{sec:Approach}

\begin{figure}[!htb]
  \center
  \includegraphics[width=0.90\columnwidth]{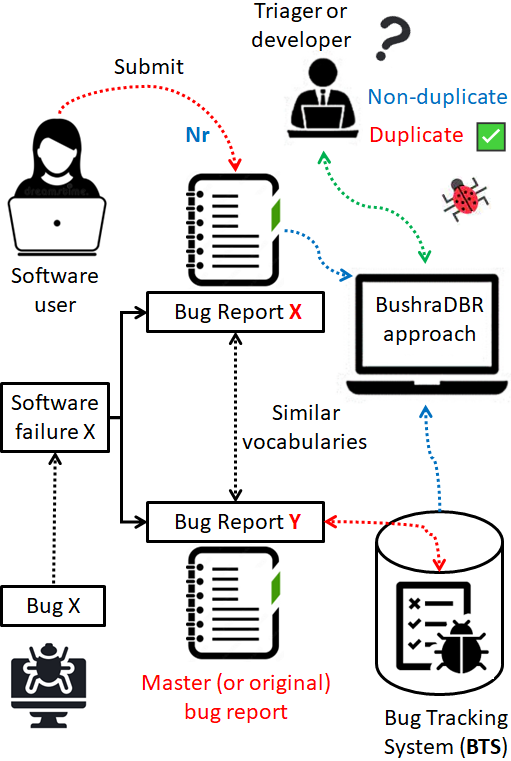}
  \caption{DBRs describe the same software failure and use similar vocabularies.}
  \label{Fig:DBRs}
\end{figure}

The main hypothesis of BushraDBR approach is that DBRs describe the same software failure and generally use similar vocabularies. Also, BushraDBR approach referred to the first reported BR for a particular bug in a software system as a master (or original) BR, while it referred to the subsequent BRs for the same bug as DBRs. Figure \ref{Fig:DBRs} shows two BRs that describe the \textit{same} software failure and use \textit{similar} vocabularies (\ie DBRs).

\begin{figure}[!htb]
	\center
	\includegraphics[width=0.95\columnwidth]{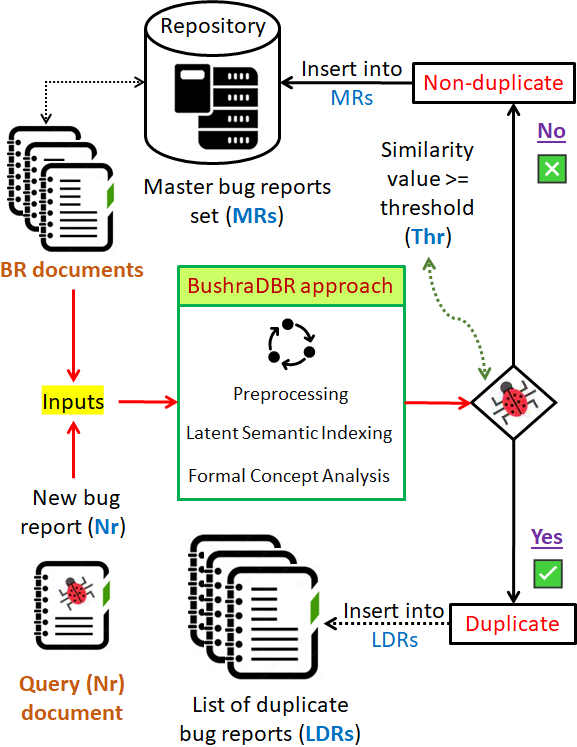}
	\caption{The DBR retrieval process - BushraDBR approach.}
	\label{Fig:Overview}
\end{figure}

The suggested approach accepts as \textit{inputs} a set (or subset) of BRs from BTS (or BRE) in addition to the newly submitted BR from the software user (\ie Nr). The proposed approach retrieves BRs that are textually similar to the newly reported BR if they exist as \textit{outputs}. The contents of each BR are textual information written in natural language (\ie BR document $\leftarrow$ summary $+$ description). An overview of BushraDBR approach is presented in Figure \ref{Fig:Overview}.

This study aims to prevent DBRs from entering the BTS. BushraDBR checks the BR and makes sure that it is not textually similar to another BR that was previously submitted to BTS. If a BR is similar to another BR, it is ignored, and if it is unique, it is added to the BTS. This process saves a lot of time, effort, and cost for the triager, as he only deals with unique BR (\ie non-duplicate). In some cases, BRs may define the same bug (or fault) with various words. In this circumstance, the suggested approach may fail to retrieve these DBRs because they use various vocabulary to describe the same bug. Table \ref{tab:DSA} displays all bug reports for DSA \cite{BushraApproach}.

\begin{table*}[!htb]
	\center
	\caption{Bug reports from the drawing shapes application data set.}
	\begin{tabular}{|c|p{0.25\textwidth}|p{0.60\textwidth}|}	\hline	
		\rowcolor{LightCyan}
		Bug ID                     &
		Short summary of the bug   &
		Bug description            \\\hline
		
		000001                                         &	
		The PDF viewer is very slow for line drawings. & 	
		"PDF is very slow to load. The text parts of the PDF appear fast, but every line drawing is very slow. While the expected result is that PDFs with line drawings should show up a lot faster without any delay." \\\hline
		000002                                                         &	
		Images of the myOval() and draw() functions are not displayed. & 	
		"The images in the draw() and myOval() functions are not displayed well. While the expected result is that for every condition in the draw() and myOval() functions, the images must be displayed well."    \\\hline	
		000003                                                               &	
		In some specific cases, the fillText function doesn't draw anything. & 	
		"In a very complex environment with a lot of sketches, the fillText method sometimes doesn't run probably and produces nothing. Thus, nothing was printed in several situations."\\\hline
		000004                      &	
		Drawing text in 2D is slow. & 	
		"The function of the 2D text doesn't efficiently scale to support drawing a large number of 2D texts on the screen at once. The performance of the DSA should be comparable to other drawing applications."\\\hline
		000005                                      &	
		Problem with the drawing of 1px-wide lines. & 	
		"The inner triangle is not 1 pixel wide and is gray rather than blue. The draw() function should draw a one-pixel-wide line."\\\hline
		000006                                                              &	
		No Scrolling of the art site contents by utilizing the mouse wheel. & 	
		"Sometimes you cannot use the mouse wheel to scroll through the art site's contents. The mouse wheel is not running well on some art sites."\\\hline\hline \rowcolor{pink!60}
		000007                                                       &	
		Incapable of using the mouse wheel to scroll on an art site. & 	
		"Scrolling the contents of the art site by clicking the scroll wheel of the mouse device does not always work well."\\\hline
	\end{tabular}
	\label{tab:DSA}
\end{table*}

According to the proposed approach, BushraDBR retrieves DBRs in three steps, as clarified in the following:

\subsection{Preprocessing of BRs} \label{sec:preprocessing}

\begin{figure}[!htb]
  \center
  \includegraphics[width=0.60\columnwidth]{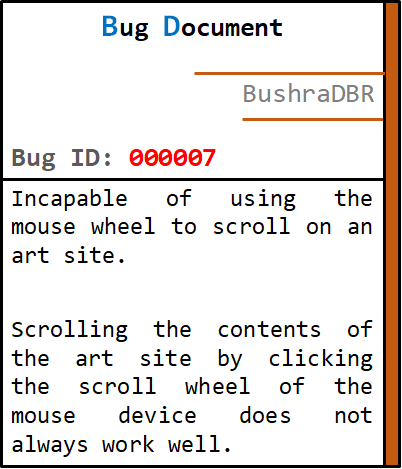}
  \caption{An example of a bug document generated by BushraDBR approach.}
  \label{Fig:BugDocument}
\end{figure}

Based on the MRs and Nr, the suggested approach generates a document for each BR. This document is named by the bug ID. The contents of each document are the bug's summary and description. Figure \ref{Fig:BugDocument} shows the bug document for the bug with ID 7 from Table \ref{tab:DSA}.

The contents of each bug document are preprocessed carefully (\cf Algorithm \ref{algo:pre-processing}), where the stop words are removed (\eg my, of, to, from, a, an, for, the, \textit{etc.}), the words of each bug document are divided based on the camel-case method (\eg fillText $\rightarrow$ fill and text), and finally every word of the document is returned to its root (\eg drawing $\rightarrow$ draw).

\begin{algorithm}[!htb]
	\caption{Preprocessing of BR documents.} \label{algo:pre-processing}
	\SetKwFunction{match}{match}
	
	\KwData{Master Bug Reports Set (MRs) and New Bug Report (Nr).}
	\KwResult{Processed BR documents (\ie query and BR documents).}
	
	// Preprocessing of all BR documents.\\
	\For{each bug report (BR) in MRs and Nr}
	{
		// Extracting BR contents.\\
		Extract the bug ID, textual summary, and description of each BR \\
		// Splitting BR words (or tokenization) via the camel-case splitting method \cite{RaFaBlasi2019}. \\
		Split the textual information of each BR into words (or tokens) \\
		// Removing English stop words from each BR by using the author's list of English stop words \cite{BushraApproach}.\\
		Remove English stop words from each BR document \\
		// Stemming BR words with WordNet \cite{GeorgeMiller95}.\\
		Return each word to its base (\ie root or stem) \\
		// The query document (\ie Nr) is named with its Bug ID.\\
		Query document $\leftarrow$ Nr document \\
		// Each BR document (\ie BR in MRs) is named with its Bug ID.\\
		BR document  $\leftarrow$ BR in MRs \\
	}
	 \Return processed documents (query and BRs)
\end{algorithm}

Algorithm \ref{algo:pre-processing} shows the step-by-step procedure for the preprocessing of BRs using BushraDBR approach. The input instances of this algorithm are all BR documents from MRs and new BR (\ie Nr), while the output instances are the processed BR documents (\ie query and BR documents).

Natural language preprocessing is employed in the majority of current studies to process the textual information of any software artifact document. BushraDBR preprocessing steps involve extracting the text from each BR, tokenization (or splitting the words of the BR), deleting (or removing) English stop words, and, at last, word stemming. Table \ref{tab:BushraPre} details the preprocessing steps of BushraDBR approach.

\begin{table*}[!htb]
 \center
 \caption{BushraDBR preprocessing steps, explanation, and a real-world example from the DSA data set.}
 \begin{tabular}{|c|p{0.13\textwidth}|p{0.42\textwidth}|p{0.28\textwidth}|}	\hline	
 \rowcolor{LightCyan}
Step ID & Step name                       &   Explanation	                    &      Example 	        \\\hline
1       & Extracting bug content          & Obtaining the text from each BR.    & 
Images of the myOval() and draw() functions are not displayed, ... \etc (\cf BR with bug ID 000002 from Table \ref{tab:DSA}) \\\hline

2       & Splitting words or tokenization & Eliminating digits (\eg 0, 5, 9), punctuation (\eg !, ?, :), and special characters (\eg @, \$, \&, \#, \%) from the text of each BR and splitting words based on the camel-case method \cite{RaFatAhmadDirasat}. & images, of, the, my, oval, and, draw, functions, are, not, displayed. \\\hline

3       & Removing stop words             & Deleting prepositions, conjunctions, and pronouns from text, such as: "a", "the", "and", "of", "are", "an", ... \etc                     & images, oval, draw, functions, displayed. \\\hline

4       & Stemming                        & Returning every word in the text to its root or stem \cite{AlkkMsiedeenSHUV16}.   &  image, oval, draw, function, display. \\\hline
    \end{tabular}
 \label{tab:BushraPre}
\end{table*}

\begin{figure}[!htb]
  \center
  \includegraphics[width=0.65\columnwidth]{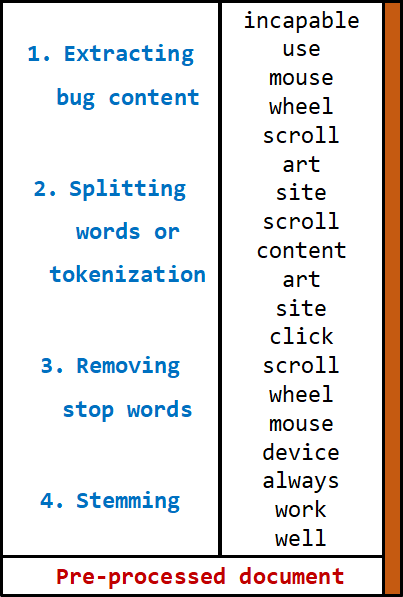}
  \caption{An example of a pre-processed bug document.}
  \label{Fig:BugID0007}
\end{figure}

Figure \ref{Fig:BugID0007} shows an example of a preprocessed bug document (\cf BR with bug ID 000007 from Table \ref{tab:DSA}). This BR document was generated by the BushraDBR approach. Only important words exist in this bug document. The preprocessing steps increase the accuracy of the LSI technique and eliminate noise.

\subsection{Measuring textual similarity between BRs using LSI} \label{sec:LSI}

\begin{figure}[!htb]
  \center
  \includegraphics[width=\columnwidth]{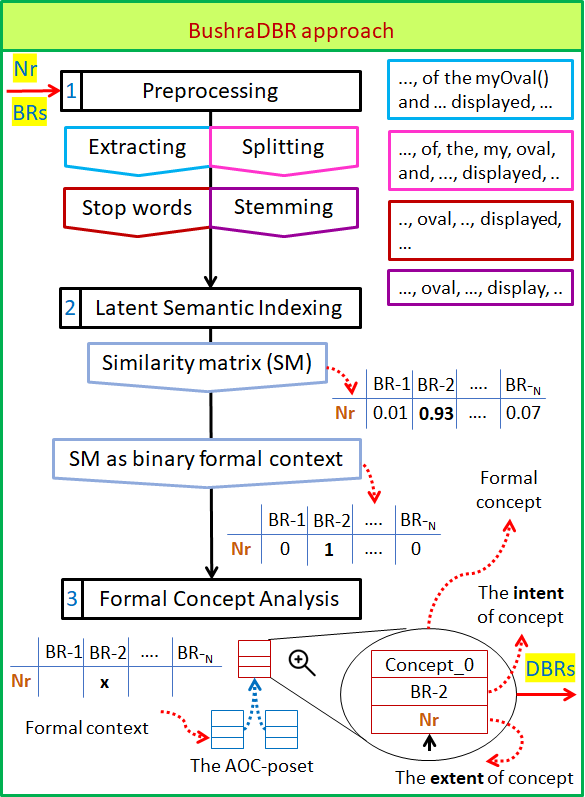}
  \caption{Retrieving DBRs using LSI and FCA techniques — BushraDBR approach.}
  \label{Fig:LSI-FCA}
\end{figure}

BushraDBR approach uses, in its core work, the LSI and FCA techniques. LSI calculates the TS scores between BRs, while FCA clusters DBRs together. Figure \ref{Fig:LSI-FCA} shows the DBR retrieval process using LSI and FCA techniques.

BushraDBR bases the retrieval of DBR on the measurement of TS between BRs. This TS measure is computed using LSI. BushraDBR depends on the truth that DBRs concerned with describing a software fault are textually closer to one another than to the remainder of BRs in the data set. To calculate TS between newly submitted BR (Nr) and BRs of the data set (MRs), BushraDBR approach goes through three steps: creating the LSI corpus (\ie BushraDBR preprocessing steps; \cf Table \ref{tab:BushraPre}); creating the Term-Document Matrix (TDM) (\resp the Term-Query Matrix (TQM)) for BRs (\resp Nr); and finally, creating the Cosine Similarity Matrix (CSM).

In order to employ LSI to find TS, BushraDBR creates a \textit{corpus} that involves a set of bug documents and query(s). In BushraDBR case, each BR in BRE represents a document, while the newly submitted BR represents a query.

TDM is of size \emph{$r \times i$}, where r is the number of terms used in the bug documents and i is the number of bug documents in the data set (or BRE). While a TQM is of size \emph{$r \times j$}, where r is the number of terms used in newly submitted bug document and j is the number of newly submitted bug documents (\ie j is equal to 1). The terms for both matrices are various because they are obtained from different bug documents (\ie Nr and BRs from BRE). Table \ref{tab:TDMDSA} shows the TDM, while Table \ref{tab:TQMEx} shows the TQM.

\begin{table}[!htb]
 \center
 \caption{TDM for bug documents from the DSA data set (partial).}
 \begin{tabular}{|c|c|c|c|c|c|c|}	\hline	
 \rowcolor{LightCyan}
	         &0001	&0002  &0003	&0004	&0005	&0006 \\\hline
method	     &0	&0	&1	&0	&0	&0 \\\hline
art	         &0	&0	&0	&0	&0	&3 \\\hline
site	     &0	&0	&0	&0	&0	&3 \\\hline
mouse	     &0	&0	&0	&0	&0	&3 \\\hline
wheel	     &0	&0	&0	&0	&0	&3 \\\hline
pixel	     &0	&0	&0	&0	&2	&0 \\\hline
function	 &0	&3	&1	&1	&1	&0 \\\hline
content	     &0	&0	&0	&0	&0	&2 \\\hline
draw	     &1	&3	&1	&3	&3	&0 \\\hline
oval	     &0	&3	&0	&0	&0	&0 \\\hline
image	     &0	&3	&0	&0	&0	&0 \\\hline
slow	     &3	&0	&0	&1	&0	&0 \\\hline
..	         &..&..	&..	&..	&..	&..\\\hline
    \end{tabular}
 \label{tab:TDMDSA}
\end{table}

BushraDBR uses the LSI technique to rank bug documents in the BTS (\ie 0001 to 0006) for the query document (\ie bug with the ID 0007), which is the new bug document (\ie Nr). BushraDBR sets the weights of terms and constructs TDM and TQM as illustrated in Tables \ref{tab:TDMDSA} and \ref{tab:TQMEx}.

\begin{table}[!htb]
 \center
 \caption{TQM for the query document (\ie Nr) from the DSA data set (partial).}
 \begin{tabular}{|c|c|c|c|c|c|c|}	\hline	
 \rowcolor{LightCyan}
      	&000007 \\\hline
art	    &2      \\\hline
wheel	&2      \\\hline
site	&2      \\\hline
use	    &1      \\\hline
scroll	&3      \\\hline
well	&1      \\\hline
content	&1      \\\hline
mouse	&2      \\\hline
..	    &..     \\\hline
    \end{tabular}
 \label{tab:TQMEx}
\end{table}

Similarity between bug documents (\ie Nr and BRs from BRE) is defined by a CSM (\cf Table \ref{tab:TexSimil}). The columns of CSM represent vectors of BRs from BRE, while the rows of CSM represent vectors of queries (\ie Nr). Equation \ref{equation:cosine} provides a Cosine Similarity (CS) that is used to calculate TS between BRs \cite{DBLPBerry} \cite{AlMsiedeenSHUVS13}.

\begin{equation}
CS (BR_q,BR_j) = \frac{\overrightarrow{BR_q}\cdot\overrightarrow{BR_j}}{\lvert \overrightarrow{BR_q \rvert} \lvert \overrightarrow{BR_j} \rvert} = \frac{\sum\limits_{i=1}^n W_{i,q} \ast W_{i,j}} {\sqrt{\sum\limits_{i=1}^n W^2_{i,q}}  \sqrt{\sum\limits_{i=1}^n W^2_{i,j}}}
\label{equation:cosine}
\end{equation}

TS between documents is calculated as a CS shown in Equation \ref{equation:cosine}, where $BR_q$ stands for the query vector and $BR_j$ for the document vector. While the $W_{i,q}$ and $W_{i,j}$ measure the weights of query and document vectors, respectively. Table \ref{tab:TexSimil} shows the calculated similarity scores between the submitted BR (\ie bug with ID 7) and the BRs from the DSA data set.

\begin{table}[!htb]
	\center
	\caption{Similarity scores (or CSM) of the DSA data set (\ie query and BR documents).}
	 \scalebox{0.80}{
		\begin{tabular}{|c|c|c|c|c|c|c|}	\hline	
			\rowcolor{LightCyan}
			& 000001	  &000002	& 000003  & 000004  &	000005  &	000006   \\\hline
			\cellcolor{GreenYellow}7 &	-0.00043  &	0.01088	& -0.03496   & 0.00317 &	0.00014 &	\cellcolor{pink!60}0.99932  \\\hline
		\end{tabular}}
	\label{tab:TexSimil}
\end{table}

Algorithm \ref{algo:MtsLSI} shows the step-by-step procedure for measuring TS between BR documents via LSI. The input instances of this algorithm are the query document (\ie Nr), BR documents, and the chosen threshold (Thr), while the output instances are the CSM and a list of DBRs (LDRs).

\begin{algorithm}[!htb]
	\caption{Measuring TS between BRs via LSI.} \label{algo:MtsLSI}
	\SetKwFunction{match}{match}
	
	\KwData{Query document (Nr), BR documents (BRs from MRs), and Threshold (Thr).}
	\KwResult{Cosine Similarity Matrix (CSM) and List of DBRs (LDRs).}
	
	CSM [$r_i$] [$c_j$] $\leftarrow$ new double [$Nr_n$][$BR_n$] \\
	LDRs $\leftarrow$ $\emptyset$ \\
	
	// Compute the TS values between the query document (Nr) and all BR documents (BRs from MRs) based on the given Threshold (Thr).\\
	\For{each query (Nr)}
	{
		// Compute the CS value between the query and each BR from the MRs.\\
		\For{each BR document}{
			Compute the CS value between the query and each BR document.\\
			CSM [$r_i$] [$c_j$] $\leftarrow$ CS (query [$BR_q$], BR document [$BR_j$]) \\
			
			\If{CS $\geq$ Thr}
			{// The query document (Nr) is considered duplicate, and the engineer (or triager) includes it in the LDRs.\\
			LDRs $\leftarrow$ Query document (Nr) \\}
			\Else
			{// The query document (Nr) is considered non-duplicate (unique BR), and the engineer (or triager) includes it in the MRs.\\
			MRs $\leftarrow$ Query document (Nr) \\}
	}} 
\Return CSM and LDRs
\end{algorithm}

\begin{figure}[!htb]
	\center
	\includegraphics[width=0.85\columnwidth]{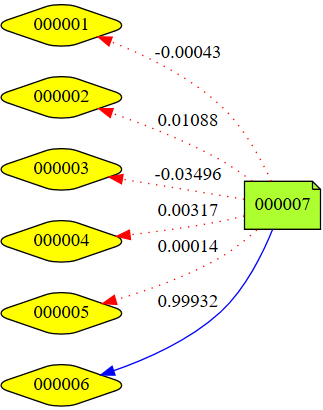}
	\caption{Textual similarity values between the query document (\ie Nr) and BRs as a directed graph.}
	\label{Fig:visualization}
\end{figure}

Figure \ref{Fig:visualization} shows the TS values between Nr and BRs from DSA as a directed graph. BushraDBR uses an external library (\ie Graphviz) to visualize the TS values between query and BR documents \cite{AlkkMsiedeenSHUV201516}. The directed graph in Figure \ref{Fig:visualization} clearly shows the retrieved DBRs by BushraDBR approach.

\subsection{Retrieving DBRs using FCA}\label{sec:LSI}

BushraDBR employs FCA to cluster similar BRs together based on LSI results (\ie the numerical CS matrix). BushraDBR transforms the CSM resulting from the LSI technique into a Binary Formal Context (BFC). Table \ref{tab:FormalCon} shows an example of a Formal Context (FC).
In this work, BushraDBR uses a standard threshold (Thr) for CS, which is equal to $0.80$ ($ CS\theta_{0.80}^{(BR_q,BR_j)}$). This implies that only pairs of BRs with a computed TS greater than or equal to the chosen threshold (\ie TS $\geq$ 0.80) are deemed DBRs. The reason (or rational explanation) behind the choice of this threshold is that DBRs generally use very similar vocabulary to describe the same bug. Also, a study of samples of DBRs available in related work (such as \cite{DBLPZhangHVIXTLJ23}, \cite{DBLP:KukkarMKNBK20}, \cite{DBLP:SunLWJK10}, and \cite{DBLPHeXY0L20}) confirms the existence of common vocabulary on a large scale among DBRs.

\begin{table}[!htb]
 \center
 \caption{A formal context of the drawing shapes application's CS matrix (\cf Table \ref{tab:TexSimil}).}
 \begin{tabular}{|c|c|c|c|c|c|c|}	\hline	
 \rowcolor{LightCyan}
	                          &0001	            &0002	        &0003	     &0004	          &0005	      &0006                 \\\hline
\cellcolor{GreenYellow}0007   &0  	  & 0      	&  0            &0               &0            &\cellcolor{pink!60}1 \\\hline
    \end{tabular}
 \label{tab:FormalCon}
\end{table}

As an instance, in the BFC of Table \ref{tab:FormalCon}, the Nr with ID $0007$ is associated with the BR document with ID $0006$ since their TS is equal to "0.99" (\cf Table \ref{tab:TexSimil}), which is higher than the chosen threshold (0.80). On the other hand, the Nr document with ID $0007$ and the BR document with ID $0002$ are not associated since their similarity is equal to "0.01", which is less than the selected threshold.
The resulting AOC-poset \cite{ALmsiedeen12} is presented in Figure \ref{Fig:AllFCAexaM}. This AOC-poset is comprised of a set of concepts. The extent and intent of some concepts link similar BR documents into one cluster (\cf Concept\_0 in Figure \ref{Fig:AllFCAexaM}).

Let's imagine that a software developer (or triager) wants to retrieve (or check) similar BRs for the newly submitted BR by a software user (\ie the bug with ID 7 from Table \ref{tab:DSA}). The BRs are created (or written) in English (\ie natural language). Each bug has a summary (or a short description) and a detailed description. Let's assume that the software developer is using the BushraDBR approach to check if a newly reported BR is a duplicate of existing BRs in the DSA data set or not. The software developer is checking similar BRs from BRE in order to retrieve possible DBRs regarding the recently submitted BR. Let's assume that the BR with ID $000007$ is the real duplicate of the BR with ID $000006$ based on the computed TS between those BRs through LSI. In this case, BushraDBR approach will classify the Nr as DBR since the suggested approach will find a lot of textually matched terms between BR documents with IDs 7 and 6. So, it is essential to consider all the textual information that is used to construct BR in BushraDBR approach.

Algorithm \ref{algo:FCA} shows the step-by-step procedure to retrieve DBRs using FCA. The input instances of this algorithm are CSM and Thr, while the output instances are ${AOC}_{BFC}$ (\ie the AOC-poset linked with $BFC$) and a list of DBRs (LDRs).

\begin{algorithm}[!htb]
	\caption{Retrieving DBRs using FCA.} \label{algo:FCA}
	\SetKwFunction{match}{match}
	// A formal context is represented by the triple F = (O, A, Br). In this triple, O is an object set (\ie Nr) and A is an attribute set (\ie BRs from MRs), while Br is a binary relation (\ie Br $\subseteq$ O × A).\\
	// BFC: a formal context, where BFC = (O, A, Br).\\
	\KwData{Cosine Similarity Matrix (CSM) and Threshold (Thr).}
	\KwResult{${AOC}_{BFC}$, $\leq_{s}$: the AOC-poset linked with $BFC$ and List of DBRs (LDRs).}
	
	BFC [$O_i$] [$A_j$] $\leftarrow$ new int [$Nr_n$][$BR_n$] \\
	LDRs $\leftarrow$ $\emptyset$ \\
    Thr $\leftarrow$ $0.80$ \\
    ${AOC}_{BFC}$ $\leftarrow$ $\emptyset$ \\
    
    // Transform the (numerical) CSM into a Binary Formal Context (BFC) based on the chosen threshold value, which is 0.80.\\
	
	\For{$O_i \gets 0$; $O_i$ $<$ $Nr_n$; $O_i\gets O_i +1$}
	{
		\For{$A_j \gets 0$; $A_j$ $<$ $BR_n$; $A_j\gets A_j +1$}{
					
			\If{CSM [$O_i$] [$A_j$] $\geq$ Thr}
			{
				BFC [$O_i$][$A_j$] $\leftarrow$ 1 \\}
			\Else
			{
				BFC [$O_i$][$A_j$] $\leftarrow$ 0 \\}
			}}

	// Apply the FCA clustering technique via the Eclipse eRCA platform \cite{eRCA}.\\
	
	Generate the AOC-poset based on BFC\\
	${AOC}_{BFC}$ $\leftarrow$ $eRCA(BFC)$ \\
    
    // The resulting AOC-poset is made up of several formal concepts (\eg concept\_0, and concept\_1).\\ 

	\For{each concept C $\in$ ${AOC}_{BFC}$}
	{// Usually, the extent and intent of the top concept (\ie $\top$) include DBRs (\cf Figure \ref{Fig:AllFCAexaM}), if any. \\
		\If{extent(C) $\neq$ $\emptyset$ and intent(C) $\neq$ $\emptyset$}
		{LDRs $\leftarrow$ extent(C) \\}
		\ElseIf{extent(C) $\neq $ $\emptyset$ and intent(C) $=$ $\emptyset$}
		{MRs $\leftarrow$ extent(C)}
		\ElseIf{extent(C) $=$ $\emptyset$ and intent(C) $\neq$ $\emptyset$}
		{intent(C) $\notin$ LDRs \\
	     intent(C) $\in$ MRs \\}
	}
	\Return ${AOC}_{BFC}$ and LDRs
\end{algorithm}

The efficiency of BushraDBR approach is assessed by its precision, recall, and F-Measure metrics \cite{Rafat2020}. For a given new BR document (\ie query or Nr), the recall metric is the ratio of rightly retrieved bug documents to the whole number of relevant BR documents, whereas the precision metric is the ratio of rightly retrieved BR documents to the complete number of retrieved BR documents. The F-Measure metric determines a trade-off among recall and precision metrics. Thus, F-Measure offers a high score just in conditions where both metrics (\ie precision and recall) are high. Recall, precision, and F-measure metrics are presented in Equations \ref{eq:recall}, \ref{eq:precision}, and \ref{eq:F-Measure}. All the evaluation metrics of the suggested approach have values between zero and one.

\begin{equation} \label{eq:recall}
Recall=\frac{\textit{relevant BR documents}\bigcap\textit{retrieved BR doc.}} {\textit{relevant BR documents}}
\end{equation}

\begin{equation} \label{eq:precision}
Precision=\frac{\textit{relevant BR doc.}\bigcap\textit{retrieved BR doc.}} {\textit{retrieved BR documents}}
\end{equation}

\begin{equation} \label{eq:F-Measure}
F-Measure= 2 \times \frac{Precision \cdot Recall}{Precision + Recall}
\end{equation}

The suggested approach relies on unstructured information (\ie bug summary and description) to retrieve DBRs from BTS. Structured information, such as component and product, does not improve the results of IR-based solutions \cite{DBLPHeXY0L20}.

\section{Experimentation}\label{sec:Exper.}

To validate BushraDBR approach, the author conducted experiments on several data sets from Bugzilla \cite{Bugzilla}. To evaluate the results, the author used recall, precision, and F-measure metrics. To implement BushraDBR approach, a system having Windows 10 Education, Intel Core i7 processor, CPU @ 2.40GHz, and 8GB RAM is used.

In this work, the Eliot \cite{Eliot} open-source data set is adopted, which is produced and available at Bugzilla \cite{EliotDataset}. Eliot is the Mozilla symbolication service, and it's part of the Tecken project \cite{Tecken}. Tecken is a project for managing and utilizing symbols at Mozilla. The Eliot product consists of two components, which are symbolication and general. The characteristics of this data set are shown in Table \ref{tab:EliotCore}.

The Firefox product \cite{Firefox} is a set of shared components utilized by the Mozilla foundation's web browser. One of these components is the WebPayments UI \cite{WebPaymentsUI}. The WebPayments UI component of Firefox product is a user interface for the WebPayments (\ie payment request API and payment handler API). The characteristics of WebPayments UI data set are given in Table \ref{tab:EliotCore}.

The Audio/Video component belongs to the core product \cite{CoreProduct}. This component deals with problems related to media (\ie video and audio) \cite{AudioVideo}. The data set of Audio/Video component consists of \textit{305} BRs.

The DOM editor component belongs to the core product \cite{CoreProduct}. This component includes bugs for the DOM editor on web pages (\ie errors with editing text on the web pages). Related HTML features for this component are: "$<$input type=text$>$, $<$textarea$>$, contenteditable, and the designMode API". The data set of the DOM editor component contains \textit{1322} BRs \cite{DOMEditor}.

The most essential LSI parameter is the number of selected term-topics (\aka the number of topics). The BushraDBR approach cannot use a constant number of topics since it deals with different data sets. Table \ref{tab:EliotCore} shows the selected number of topics (\ie $K$) for each data set.

The reader can download all BRs of any data set from Bugzilla as an XML file. BushraDBR approach generates all BR documents based on this XML file. BushraDBR names each BR document by the same bug ID in the XML file.

\begin{table*}[!htb]
 \center
 \caption{Characteristics of the data sets that are used in experiments.}
 \begin{tabular}{|c|c|c|c|c|c|}	\hline	
 \rowcolor{LightCyan}
ID & Product name              & Component (\ie Data set)  & Total number of BRs & Number of DBRs & $K$   \\\hline
1  & Drawing shapes application& DSA                       & 0007                & 	001	         & 006   \\\hline
2  & Eliot                     & Symbolication and General & 0017                & 	001	         & 016  \\\hline
3  & Firefox                   & WebPayments UI            & 0126                & 	001	         & 125 \\\hline
4  & Core                      & Audio/Video               & 0305                & 	001	         & 304 \\\hline
5  & Core                      & DOM: Editor               & 1322                & 	001	         & 030  \\\hline
    \end{tabular}
 \label{tab:EliotCore}
\end{table*}

In this study, a threshold of $0.80$ (\ie Thr $\leftarrow$ $0.80$) was chosen to retrieve DBRs. This means that BRs with a TS value equal to or greater than $0.80$ are considered DBRs. The reason for choosing this threshold is that BRs that use similar vocabulary with a percentage greater than or equal to $0.80$ are mostly duplicates. In fact, it is not possible for a developer to find two reports that use exactly the same vocabulary (100\%). But the developer can find two reports that use common vocabulary in a large ratio. BRs are usually submitted by different users to BTS. Each user uses different vocabulary to describe the submitted BR. If users submit many BRs to describe the same error, they will most likely use similar vocabulary to describe it.

Experiments show the ability of BushraDBR approach to retrieve DBRs in an efficient and accurate manner. For instance, results show that BRs with IDs \textit{1819152} and \textit{1819151} from the Eliot data set are DBRs. In this case, the BR with ID \textit{1819152} (\ie Nr) is ignored and included in the LDRs. Thus, BushraDBR approach saves the engineer's time and effort, as there is no need to address this duplicate BR. Table \ref{tab:EliotTS} shows the TS obtained from the Eliot data set.

\begin{table*}[!htb]
	\center
	\caption{Cosine similarity matrix for the Eliot data set.}
		\begin{tabular}{|c|c|c|c|c|c|c|c|c|c|c|}	\hline	
		\rowcolor{LightCyan}	
		&1475334 &	1649535 &	1707879 &	1729698 &	1741434 &	1745533	& 1768863	& 1801169 &	1801212 \\\hline
		\cellcolor{GreenYellow}001819152  &	-0.00294&-0.00188 &0.00907 &0.01036 &	0.04272 & -0.00197&	0.00670	 &-0.01518&	0.01662 \\\hline\hline
		\rowcolor{LightCyan}
		&1811236  &1811299	&1812345&	1814509	&1815981&	1815982	&1819151&&\\\hline
		\cellcolor{GreenYellow}001819152&	-0.00454&	-0.00030&	0.02736&	-0.01092&	0.00708	&-0.00400&	\cellcolor{pink!60}0.99822&&\\\hline
		\end{tabular}
	 \label{tab:EliotTS}
\end{table*}

As an illustrative example, let's say that a user has reported a BR regarding the WebPayments UI component, which belongs to the Firefox browser, by using Bugzilla BTS. The user has pressed the link to insert a new BR. A user added a summary for the BR as follows: "Need animation on the pay submit button". Then he inserted a description to this BR as follows: "As shown in UX specs, once users click on the pay submit button, we need to have an animation to show order payment sent to merchant". Finally, he pressed the submit BR button. By using BushraDBR, the approach finds that the newly added BR with ID \textit{1510066} is very similar to the BR with ID \textit{1490824} from the WebPayments UI data set (\cf Table \ref{tab:WebPayUI}). In this case, the newly submitted BR is ignored and included in the LDRs.

\begin{table*}[!htb]
	\center
	\caption{Cosine similarity matrix for the WebPayments UI data set (partial).}
		\begin{tabular}{|c|c|c|c|c|c|c|c|c|c|c|}	\hline	
			\rowcolor{LightCyan}	
	&1427949&	1427953	&1432909&	1432940	&1490824&	1432943	&1432945&	1432958&	1435114&..\\\hline
	\cellcolor{GreenYellow}1510066	&0.01822&	0.00348	&0.02232&	0.05142	&\cellcolor{pink!60}0.92465&	0.00149	&0.00331&	0.01348	&0.00261&..\\\hline
		\end{tabular}
	\label{tab:WebPayUI}
\end{table*}

A complete tutorial showing how to implement BushraDBR on the audio-video data set is available on BushraDBR web page \cite{BushraApproach}. All steps involved in the approach are fully illustrated. The purpose of this tutorial is to show how the implementation of BushraDBR works. Table \ref{tab:Au/Veo} shows the TS of the Audio/Video data set.

\begin{table*}[!htb]
	\center
	\caption{Cosine similarity matrix for the Audio/Video data set (partial).}
		\begin{tabular}{|c|c|c|c|c|c|c|c|c|c|c|}	\hline	
			\rowcolor{LightCyan}	
			       & 1368902	&1532646   &	1545235	                &1545847	& 1561960	& 1565493 &	1571258	& 1571583 &	1571912 &..\\\hline	
			\cellcolor{GreenYellow}1545237& 0.00107	&0.00186   &\cellcolor{pink!60}0.84855	&0.00752    &	0.00270	& 0.00391 & 0.00202	& 0.00636 &	0.00293 &..\\\hline		
		\end{tabular}
	\label{tab:Au/Veo}
\end{table*}

The key limitation of utilizing FCA as a clustering technique in BushraDBR is that FCA works only with BFC (\ie 1 or 0). Where BushraDBR approach considers that the similarity value of 0.97 (\resp 0.79) is the same as 0.80 (\resp 0.05). Table \ref{tab:DOMEdi} shows the TS of the DOM editor data set.

\begin{table*}[!htb]
	\center
	\caption{Cosine similarity matrix for the DOM editor data set (partial).}
		\begin{tabular}{|c|c|c|c|c|c|c|c|c|c|c|}	\hline	
			\rowcolor{LightCyan}			
			&1005776 &362694	&176525	   &1311623&	1043099       &666037  &1016372 & 229572 &	579763& ..\\\hline
			\cellcolor{GreenYellow}001841744	 &0.05002	& 0.07664 &\cellcolor{pink!60}0.96201 &0.05911 &0.00183 & 0.07517&	0.01558 &	0.08070	 & 0.01767 & ..\\\hline	
		\end{tabular}
	\label{tab:DOMEdi}
\end{table*}

Also, results show that BushraDBR approach is able to retrieve DBRs when these BRs share a common vocabulary. Otherwise, the approach will fail to retrieve DBRs. When the users use different vocabulary to describe the same bug, BushraDBR will be unable to retrieve DBRs. This means that the LSI technique may not be dependable (or should be improved with ML methods) in all cases to retrieve DBRs. For example, if we have two BRs describing the same bug and the description of the first is written as "doesn't run in dialogues", while the description of the second is written as "no longer runs as anticipated". In this case, both descriptions have the same meaning but use different vocabulary. BushraDBR will consider both reports as original BRs. 

\begin{figure}[!htb]
	\center
	\includegraphics[width=\columnwidth]{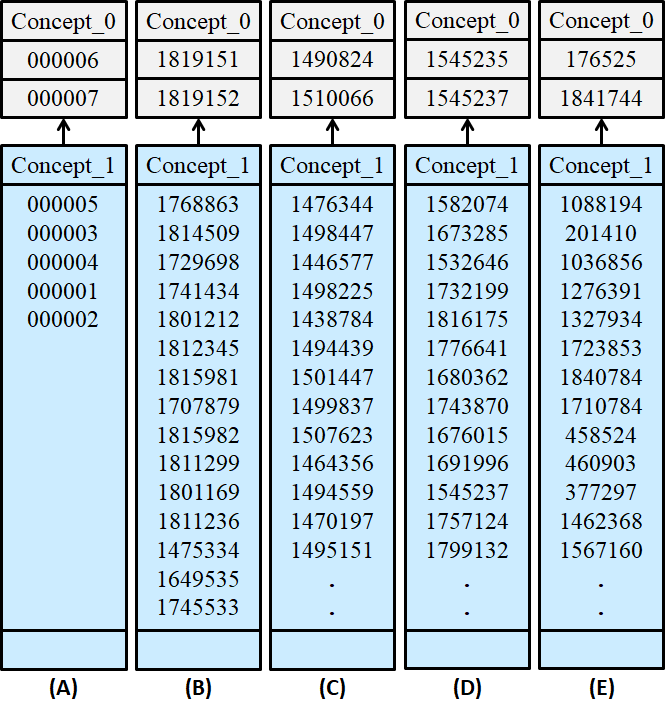}
	\caption{The AOC-poset for each data set in the experiments: (\textbf{A}) DSA, (\textbf{B}) Eliot, (\textbf{C}) WebPayments UI [\underline{partial}], (\textbf{D}) Audio/Video [\underline{partial}], and (\textbf{E}) DOM editor [\underline{partial}].}
	\label{Fig:AllFCAexaM}
\end{figure}

Figure \ref{Fig:AllFCAexaM} shows the AOC-poset for each data set. The AOC-poset of Figure \ref{Fig:AllFCAexaM} shows two formal concepts for each data set (\ie Concept\_0 and Concept\_1). In general, the top concept (\ie $\top$) contains DBRs, if any. For example, the extent and intent of Concept\_0 include similar BRs (\ie DBRs). Where the extent of Concept\_0 contains the query document (\ie Nr), while the intent of Concept\_0 represents a similar BR from BRE.

Table \ref{tab:Evaluationmetrics} shows the evaluation metrics for BushraDBR's results from different data sets. Considering the recall metric, its value is 1 (or 100\%) for all retrieved DBRs from the different data sets. This implies that all DBRs are correctly retrieved. Also, the precision metric value is high (100\%) for all retrieved DBRs. This means that all retrieved DBRs are relevant. The F-measure value depends on the values of recall and precision metrics; thus, its value is 1 for all retrieved DBRs.

\begin{table}[!htb]
	\center
	\caption{Evaluation metrics for BushraDBR results.}
		\begin{tabular}{|c|c|c|c|c|}	\hline	
			\rowcolor{LightCyan}	
			ID & Data set      & Recall	   & Precision	  & F-measure \\\hline
			01 & DSA           &100\%      &100\%         &100\%      \\\hline	
			02 & Eliot         &100\%      &100\%         &100\%      \\\hline	
			03 & WebPayments.. &100\%      &100\%         &100\%      \\\hline	
			04 & Audio/Video   &100\%      &100\%         &100\%      \\\hline	
			05 & DOM: editor   &100\%      &100\%         &100\%      \\\hline	
		\end{tabular}
	\label{tab:Evaluationmetrics}
\end{table}

In the audio/video data set, BushraDBR retrieves the report with ID \textit{1545237} as a duplicate of the report with ID \textit{1545235}. The author performed a manual check of the content of each BR. The author found that the summary and description of both reports are textually similar to each other. Where the TS value between the two reports is equal to \textit{0.84855} (\cf Table \ref{tab:Au/Veo}). After conducting this manual evaluation of BushraDBR results, the evaluation metrics (\cf Table \ref{tab:Evaluationmetrics}) proved the accuracy, strength, and solidity of BushraDBR to retrieve DBRs. Thus, BushraDBR is able to prevent DBRs using LSI and FCA.

BushraDBR approach can be used only to retrieve DBRs from BTS based on the recently submitted BR. In this work, BushraDBR uses textual information to retrieve DBRs; thus, BRs vocabulary is very sensitive to BushraDBR (\ie a threat to internal validity). Hence, based on the vocabulary used, BushraDBR may succeed or fail. In this study, the author validates his approach using different data sets from Bugzilla. This could pose a challenge to applying the suggested approach to another BTS in general (\ie a threat to external validity). However, the considered data sets are popular and cover different sizes. BushraDBR assumes that the user has constructed the submitted BR correctly in terms of the report summary and description. Sometimes it is possible that the BRs used are incorrectly constructed or that some parts of the BR are missing (\eg report without description), which may cause the proposed approach to be inaccurate (\ie a threat to construct validity).

\begin{figure}[!htb]
	\center
	\includegraphics[width=\columnwidth]{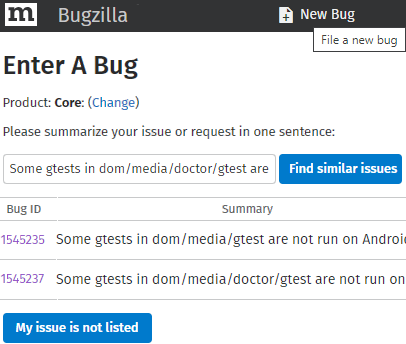}
	\caption{An example of the DBR retrieval feature in Bugzilla based on the bug summary.}
	\label{Fig:Bugzilla}
\end{figure}

Bugzilla \cite{Bugzilla} offers a feature that suggests similar BRs when a user submits a new BR (\cf Figure \ref{Fig:Bugzilla}). The current implementation of this feature is only based on text in the summary part of BRs and does not take into account the description part of BRs. Also, the implementation does not count the frequency of words that arise in BRs. Thus, BushraDBR tool would retrieve DBRs better by using both the summary and description parts of BRs.
Mozilla utilizes Bugzilla as its BTS, and Bugzilla implements an important feature to retrieve DBRs called Full-Text Search (FTS) \cite{DBLPZhangHVIXTLJ23}. In general, this feature relies on a BRs database (BRE) and issues SQL queries to search in the BRs database \cite{JournalAlMsiedeenDB}.

In order to retrieve DBRs via BushraDBR approach, the initial prototype was developed and accessible through the BushraDBR web page \cite{BushraApproach}. To extract the report summary and description, the author has established an XML parser for this target based on the Jdom library \cite{jdom}. Jdom is an open-source Java-based solution for reading and manipulating XML files. In order to use LSI, the author has built an LSI implementation available on BushraDBR web page based on the JAMA package \cite{JAMA}. JAMA is a Java library for algebra, and it is able to find the Singular Value Decomposition (SVD) for specific documents belonging to a single corpus. For applying FCA, the author utilized the eRCA tool \cite{eRCA}. Finally, to visualize the AOC-poset (\resp similarity scores between BRs), BushraDBR employs the Graphviz library \cite{AlkkMsiedeenSHUV201516} in its implementation.

\section{Conclusion and future work}\label{sec:conclusion}

This paper has introduced a novel approach called BushraDBR targeted at automatically retrieving DBRs using LSI and FCA. BushraDBR aimed to prevent developers from wasting their resources, such as effort and time, on previously submitted BRs. The novelty of BushraDBR is that it exploits textual data in BRs to apply LSI and FCA techniques in an efficient way to retrieve DBRs.
BushraDBR prevents DBRs before they occur by comparing the newly reported BR with the rest of the BRs in the repository. The suggested approach had been validated and evaluated on different data sets from Bugzilla. Experiments show the capacity of BushraDBR approach to retrieve DBRs in an efficient and accurate manner.
Regarding BushraDBR's future work, the author plans to extend the current approach by developing an ML-based solution to retrieve DBRs and prevent duplicates before they start. Also, he plans to compare BushraDBR (\ie an IR-based approach) with current ML-based approaches. Furthermore, additional empirical tests can be conducted to verify BushraDBR approach using open-source and industrial data sets. There is also a necessary need to conduct a \textit{comprehensive survey} and make comparisons between all current approaches relevant to DBR retrieval.

%% file: ijcds_biographies.tex
\vspace{-.5cm}
\begin{IEEEbiography}[{\includegraphics[width=1in,height=1.3in,clip,keepaspectratio]{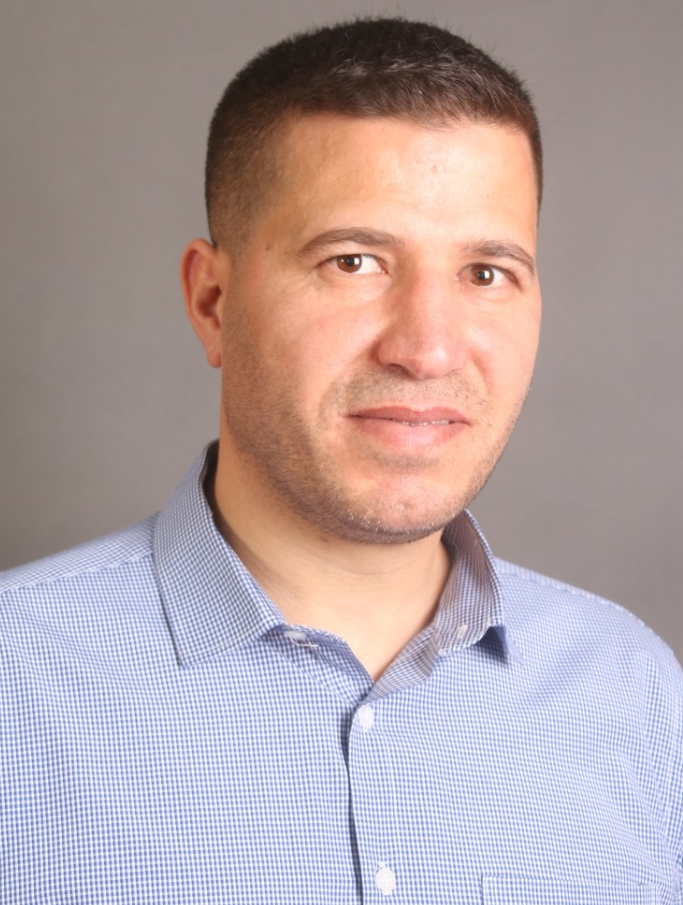}\squeezeup}]
{Ra'Fat Al-Msie'Deen} is an Associate Professor in the Software Engineering department at Mutah University since 2014. He received his PhD in Software Engineering from the Université de Montpellier, Montpellier - France, in 2014. He received his MSc in Information Technology from the University Utara Malaysia, Kedah - Malaysia, in 2009. He got his BSc in Computer Science from Al-Hussein Bin Talal University, Ma'an - Jordan, in 2007. His research interests include software engineering, requirements engineering, software product line engineering, feature identification, word clouds, and formal concept analysis. Dr. Al-Msie'Deen aimed to utilize his background and skills in the academic and professional fields to enhance students expertise in developing software systems. Contact him at \href{mailto:rafatalmsiedeen@mutah.edu.jo}{\faEnvelope}\href{mailto:rafatalmsiedeen@mutah.edu.jo}{ rafatalmsiedeen@mutah.edu.jo}. Also, you can reach him using different alternatives: \href{https://rafat66.github.io/Al-Msie-Deen/}{\faGithub}\href{https://rafat66.github.io/Al-Msie-Deen/}{ author's page @ github.io}, \href{https://www.linkedin.com/in/ra-fat-al-msie-deen-08895062/}{\faLinkedin}\href{https://www.linkedin.com/in/ra-fat-al-msie-deen-08895062/}{ LinkedIn}, \href{https://www.researchgate.net/profile/Rafat-Al-Msiedeen}{\includegraphics[width=0.022\textwidth]{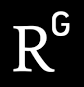}}\href{https://www.researchgate.net/profile/Rafat-Al-Msiedeen}{ ResearchGate}, or \href{https://orcid.org/0000-0002-9559-2293}{\includegraphics[width=0.022\textwidth]{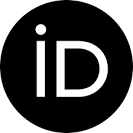}}\href{https://orcid.org/0000-0002-9559-2293}{ Orcid}.
\end{IEEEbiography}